\documentclass[debug,overfull]{epl}
\usepackage{epsfig,amssymb,amsmath,hyperref,bm}
\usepackage{graphics,graphicx,psfrag}
\def\be{\begin{equation}}
\def\ee{\end{equation}}
\def\beq{\begin{eqnarray}}
\def\eeq{\end{eqnarray}}

\newcommand\tJp{\tilde J_{\perp}}
\newcommand\tJ{\tilde J}

\title{On charge and spin ordering in a one-dimensional model with frustrating 
interactions.}
\shorttitle{Charge and spin ordering in a one-dimensional model}
\author{M. S. Laad\inst{1}\thanks{E-mail: \email{M.S.Laad@lboro.ac.uk}} 
\and S. Lal\inst{2}\thanks{E-mail: \email{sanjayl@thp.uni-koeln.de}}}

\institute{                    
  \inst{1} Department of Physics, Loughborough University, LE11 3TU, UK\\
  \inst{2} Institut f\"ur Theoretische Physik, 77 Z\"ulpicher Strasse, 50937, 
K\"oln, Germany
}

\begin{document}

\maketitle

\begin{abstract}
We study a one-dimensional extended Hubbard model with longer-range 
Coulomb interactions at quarter-filling in the strong coupling limit.  
We find two different charge-ordered ground states as the strength 
of the longer range interactions is varied. At lower energies, these 
CO states drive two different spin-ordered ground states. A variety 
of response functions computed here bear a remarkable resemblance to 
recent experimental observations for organic TMTSF systems, and so 
we propose that these systems are proximate to a QCP associated with 
$T=0$ charge order. For a ladder system relevant to $Sr_{14}Cu_{24}O_{41}$, 
we find in-chain CO, rung-dimer, and orbital antiferromagnetic 
ordered phases with varying interchain couplings and superconductivity 
with hole-doping.
\end{abstract}

%
%

Electron crystallization, or charge ordering (CO) due to 
interactions, is an issue of enduring interest in condensed matter 
physics.  The study of the conditions favoring CO, 
along with its competition with metallic and/or superconducting 
states constitutes a problem of wide-ranging interest for a host 
of real systems~\cite{org} composed of weakly coupled single 
chains/ladders.

In this work, we study this issue within an extended quarter-filled 
Hubbard model on a linear chain, described by,
\be
H_{\mathrm{eff}} = -t\sum_{i,\sigma}(C_{i\sigma}^{\dag}C_{i+1,\sigma}+h.c) 
+ (U-2zP)\hspace*{-0.1cm}\sum_{i}n_{i\uparrow}n_{i\downarrow}
+ V\sum_{i}n_{i}n_{i+1}
+ P\sum_{i}n_{i}n_{i+2}
\ee

 In one-dimension, the
spin fluctuations are those of an ideal $S=1/2$ XXX AF chain, while the charge
fluctuations are described by the Hamiltonian
\be
H_{\mathrm{c}}=-t\sum_{i}(c_{i}^{\dag}c_{i+1}+h.c) +
(V-J/4)\sum_{i}n_{i}n_{i+1} + P\sum_{i}n_{i}n_{i+2}
\ee
that describes a model with frustrating interactions.
In 1D, the projected fermions are spinless fermions with a hard-
core constraint.  This model has been considered as a model for studying the
effect of frustration on electron crystallization~\cite{kats}. 
For narrow-band systems, we consider the limit $t<<(V-J/4),P$.
In this regime, we employ 
an extension of the trick used for the $1d$ next-nearest neighbor Ising chain:
for $t=0$, we notice that with $(V-J/4)>2P$, the ground state is the usual CDW
(Wigner) crystal for $n=1$.  With $2P>(V-J/4)$, however, the dimerized state
(Peierls) is the ground state, written schematically as (11001100....). 
Splitting this in a slightly different way, we have [...(01)(10)(01)(10)...].
Associating a pseudospin $\tau=1/2$ operator, with $\tau^{z}=+1$ for (10) and 
-1 for (01), the state is antiferromagnetic and doubly degenerate in terms of
the $\tau_{i}^{z}$.  For small $t$, this is an attractive trick because (in
spin language) the transverse term does flip the $\tau_{i}^{z}$, but cannot 
break a pair.  So one obtains,
\be
H_{\mathrm{eff}}=-\sum_{l}[2t\tau_{l}^{x}+(V-J/4-2P)\tau_{l}^{z}\tau_{l+1}^{z}]
\ee
  
  This is just the Ising model in a transverse field, which has been studied
extensively in 1D.  If $(V-J/4-2P)<0$, the ground state is 
ferromagnetically ordered in $\tau^{z}$, i.e, it corresponds to a Wigner CDW.
 For $(V-J/4-2P)>0$, the Peierls dimer order results in the ground state.  At 
$(V-J/4-2P)<2t$, the quantum disordered phase
has short-ranged pseudospin correlations, and is a charge
``valence-bond" liquid.  
The quantum critical point at $(V-J/4-2P)=2t$ separating these phases is a
deconfined phase with gapless pseudospin ($\tau$) excitations, and 
power-law fall-off in 
the pseudospin-pseudospin correlation functions.  Correspondingly, 
the density-density 
correlation function has a power-law singular behavior at low energy, 
with an exponent
$\alpha=1/4$ characteristic of the $2D$ Ising model at criticality.  
For $P=0$, the metallic phase for $V \le 2t$ is a Luttinger liquid,
and in this limit, the low-energy physics is qualitatively similar to that of 
the usual $t-J$ model.  The ``Mott" insulating state for $V > 2t$
 has Wigner CO in the 
ground state, and the M-I transition is of the Kosterlitz-Thouless
 type~\cite{tsvelik}.
 
  The full Hamiltonian in our case for the strong-coupling limit is now 
given by
\be
H_{\mathrm{eff}}=\hspace*{-0.1cm}-\hspace*{-0.1cm}\sum_{l}[2t\tau_{l}^{x}+(V\hspace*{-0.1cm}-\hspace*{-0.1cm}\frac{J}{4}\hspace*{-0.1cm}-\hspace*{-0.1cm}2P)\tau_{l}^{z}\tau_{l+1}^{z}]
+ J\sum_{l}{\bf S_{l}}.{\bf S_{l+1}}  
\ee

  To study the magnetic phases, we adapt the Ogata-Shiba~\cite{ogata}
technique for our case.  This is possible if $J<<t,V$, in which case, the 
pseudospin part is first solved exactly (this is possible because of the 
known exact solution of the 1D transverse field Ising model), and the exchange
part is then treated as a perturbation.  Writing the total wavefunction as a 
product of a spin and pseudospin wavefunction (where the spin wavefunction is
defined in a Hilbert space of dimension $2^{N}$), 
i.e, $|\psi>=|\tau>\otimes|{S}>$,
and following standard degenerate perturbation theory, the spin degeneracy is 
lifted by the correction (of order $1/L$):
\beq
<H_{\mathrm{eff}}>' = -2t<\tau^{x}>
+ \sum_{l} J_{l,l+1}({\bf S_{l}}.{\bf S_{l+1}}-1/4)
\eeq
 where the average $<..>'$ denotes that the average is taken over the exact
ground state $|\tau>$ of the pseudospin part above, i.e. $<A>'=<\tau|A|\tau>$ 
and $J_{l,l+1}=\frac{1}{N}(V-\frac{J}{4}-2P)<\tau_{l}^{z}\tau_{l+1}^{z}>$. 

An interesting fact now emerges: Wigner CO (FM order of $\tau$) results in 
an HAFM $S=1/2$ spin model with the Hamiltonian
$H_{\mathrm{s}}=J\sum_{i}{\bf S}_{i}.{\bf S}_{i+1}$. This gives rise to a 
gapless AF ground
state for the spin degrees of freedom.  The charge (holon) 
excitations are gapped; this corresponds to a linear confining potential for
holons.  On the other hand, Peierls dimerization in the
charge sector (AF Neel order of $\tau$) gives rise to dimerization in the 
spin sector, with the Hamiltonian
$H_{\mathrm{s}}=J\sum_{i}[1+(-1)^{i}\delta]{\bf S}_{i}.{\bf S}_{i+1}$.

  Translated into fermion variables, this yields a sine-Gordon problem with 
$\beta^{2}=2\pi$, and describes an instability to a {\it singlet} pinned
ground state commensurate with the Peierls CO setting in at higher energies.    
The elementary excitations are solitons carrying $S^{z}=\pm 1$.
Scaling theory predicts a dimer gap, 
$\Delta_{\mathrm{d}} \simeq \delta^{2/3}$.  
Exactly at $\beta^{2}=2\pi$, the SG model has just {\it two} $S^{z}=0$
breather excitations with opposite parity~\cite{haldane}, the lowest, even 
parity breather being degenerate with the $S^{z}=\pm 1$ soliton doublet,
forming a $S=1$ triplet, while the second odd-parity breather is a singlet 
with a gap, $\sqrt{3}\Delta_{\mathrm{d}}$.
  It is important to notice that {\it both} charge and spin order
arise from long range Coulomb interactions, and do not involve an electron
phonon coupling mechanism.  

  Let us consider the implications of having the CO state in the high-$T$ 
regime, where one could imagine the system to be effectively one-dimensional.
In particular, we want to look at the $\omega,T$ dependence of the various 
response functions at high-$T$.  Using the exact 
solution of the pseudospin model in 1D, the high $T$ (in the ``quantum critical"
regime) behavior can be explicitly derived~\cite{sachdev}.
In fact, near Ising criticality, the response function, $\chi(r)\simeq r^{-1/4}$
 where $r=(x^{2}+\tau^{2})^{1/2}$ (with the velocity $v$ set to unity).  This
relation is still valid away from criticality in the ``short range" region,
$r<<\Delta_{\tau}$, where $\Delta_{\tau}$ is the pseudospin (in our case charge
gap) gap of the 1D-TFIM.  Using this asymptotic form, we have
\be
\chi_{\mathrm{crit}}(0,\omega)=-\frac{\sin(2\pi\Delta)}{(2\pi T)^{2-4\Delta}}B^{2}(\Delta-iS,1-2\Delta)
\ee
where $S=\frac{\omega}{4\pi T}$, and $\Delta=1/16$ is the conformal dimension.
$B(x,y)$ is the beta function.

In the quantum critical region, an illuminating form is
\be
\chi(k,\omega)=\frac{\chi(0,0)}{1-i\omega/\Gamma_{\mathrm{R}}+k^{2}\xi^{2}-(\omega/\omega_{1})^{2}}
\ee
where $\Gamma_{\mathrm{R}}=(2\tan(\pi/16)k_{\mathrm{B}}T/\hbar)e^{-\Delta_{\tau}/k_{\mathrm{B}}T}$, $\omega_{1}=0.795(k_{\mathrm{B}}T/\hbar)$ and 
$\xi=\hspace*{-0.05cm}1.28(c\hbar/k_{\mathrm{B}}T)e^{\Delta_{\tau}/k_{\mathrm{B}}T}$,
 are determined solely by $T$ and the fundamental natural constants, as expected in the QC regime.  Here, $\Delta_{\tau}$ is the energy gap to charge excitations in the Wigner/Peierls CO states described above.
 This represents the collective charge susceptibility, and the optical conductivity follows directly from 
 $\sigma(\omega)=-i\omega\chi(0,\omega)$, giving,
\be
\sigma(\omega)=\frac{\chi(0,0)}{\Gamma_{R}}\frac{\omega^{2}}{(1-\omega^{2}/\omega_{1}^{2})^{2}+(\omega/\Gamma_{\mathrm{R}})^{2}}
\ee

The corresponding frequency-dependent dielectric function is obtained from
$\epsilon(\omega)=1+(4\pi i\sigma(\omega)/\omega)$, and the electronic 
contribution to the Raman scattering is estimated therefrom to be given by
$I_{\mathrm{R}}(\omega)=\mathrm{Im}(1/\epsilon(0,\omega))$, for light polarized along the chain 
axis.  
In terms of the charge susceptibility, this is simply,
\be
I_{\mathrm{R}}(\omega)=\mathrm{Im}\frac{1}{\epsilon(\omega)}=\frac{\frac{4\pi\chi(0,0)}{\Gamma_{\mathrm{R}}}F(\omega,T)}{1+(\frac{4\pi\chi(0,0)}{\Gamma_{\mathrm{R}}})^{2}F^{2}(\omega,T)}
\ee
where 
$F(\omega,T)=\frac{\omega}{(1-\omega^{2}/\omega_{1}^{2})^{2}+(\omega/\Gamma_{\mathrm{R}})^{2}}$.

$\chi"(k,\omega)$ has its maximum value at $\omega_{m}=\omega_{1}-i(\omega_{1}^{2}/\Gamma_{\mathrm{R}})$, implying that the collective mode broadens and 
shifts to higher energy {\it linearly} in $T$ with increasing $T$ at high 
temperatures.  
 Further, the $T$-dependent damping rate of the 
collective mode correlates well with the relaxational peak seen in transport,
underlying their common origin.  In fact, the $dc$ resistivity is linear in $T$
at high $T$, with ``insulating" features showing up at lower $T$.  In our picture, these are collective
(longitudinal) bosonic
 charge-density modes in the high-$T$ quantum critical region
above an incipient CO transition (expected to occur at low $T$).  In fig.(1), 
we show the electronic Raman lineshape as a function of $\omega/T$.  The sharp
low energy peak corresponds to the collective charge density fluctuation mode
of the CO ground state.   
\begin{figure}[htb]
\begin{center}
\scalebox{0.8}{
\includegraphics{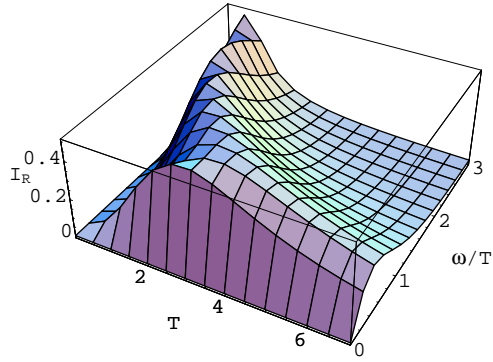}
}
\end{center}
\caption{A three-dimensional plot of the Raman intensity $I_{R}$ versus a 
scaled energy $\omega/T$ and temperature $T$ for parameter values of the 
original model which constitute a gap $\Delta_{\tau}=0.05 k_{\mathrm{B}}$.}
\label{raman}
\end{figure}

  The corresponding frequency-dependent dielectric constant also shows an 
explicit $\omega/T$ scaling in the QC regime, or generally, at high-$T$,  
it shows strong $T$-dependence.  From fig.(2), we see 
that it becomes $\omega$-independent at high $T$, but appreciably increases as 
$T$ is lowered, with a maximum at $\omega\simeq T$. 
\begin{figure}[htb]
\begin{center}
\scalebox{0.8}{
\includegraphics{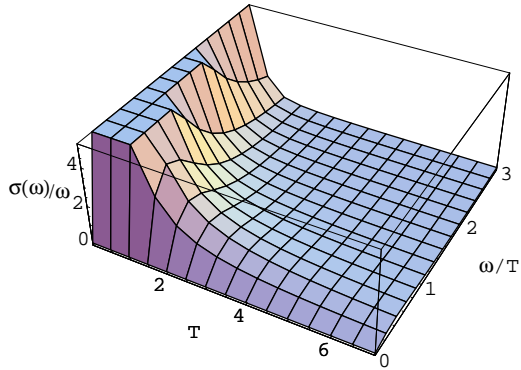}
}
\end{center}
\caption{A three-dimensional plot of $\sigma(\omega)/\omega$ versus a 
scaled energy $\omega/T$ and temperature $T$ for 
$\Delta_{\tau}=0.05 k_{\mathrm{B}}$. The limiting behaviors are:
(1) $\sigma(\omega) \simeq \frac{\omega^{2}}{T^{11/4}}$ for high $T$,
and $\sigma(\omega) \simeq \frac{1}{\omega^{3/4}}$ for low $T$.  
(2) $\epsilon'(\omega,T)=1+\frac{const}{T^{7/4}}$ for high $T$, 
and $\epsilon'(\omega,T) \simeq \frac{-1}{\omega^{7/4}}$ for low $T$. 
}
\label{sigmaom}
\end{figure}

  The fact that organic charge transfer salts~\cite{bra} exhibit features very similar to those found 
above has interesting implications.  In light of our results, these anomalous features can now be identified with proximity to an underlying 
quantum critical point associated with charge (Wigner/Peierls) ordering. 
We recall that very recent work~\cite{mon,bishop} shows that the dimerized insulating state in TMTSF systems has charge order at low $T$.
Interestingly, $\epsilon'(0,\omega)$ indeed shows appreciable increase as $T$ is lowered, further supporting an interpretation based on
proximity to an underlying CO ground state~\cite{bra}.
 Hence, we conclude that observation of these features in
TMTSF systems constitutes strong evidence that this system is close to a putative QCP associated with charge
order.  Observation of dimerized/Neel ordered AFM states co-existing CO states at low $T$ are also naturally understood in light of 
the analysis above~\cite{bishop}.      

{\bf Two-chain Ladders}
We now consider the strong coupling version of a coupled two-chain ladder system, with
each chain being described by $H$ as in eq.(1).  In the strong coupling limit, where each
chain is described by a TFIM for charge degrees of freedom, the coupled chain model is 
constructed as follows.  
For $U\rightarrow\infty$, and $V,P>t$ (but $(V-2P)$ comparable to $t$),
the charge degrees of freedom
of the fermionic problem for each chain are described by an effective
pseudospin model on n-n bonds, via the effective Hamiltonian,
\be
H^{chain} = -\sum_{j} [2t\tau_{j}^{x}
+ (V - J/4 - 2P)~\tau_{j}^{z}\tau_{j+1}^{z}]
\label{onechain}
\ee
Rotating the pseudospin axis such that $\tau^{x}\rightarrow\tau^{z}$,
$\tau^{z}\rightarrow - \tau^{x}$ and coupling two such chains via
an interaction coupling $U_{\perp}$
and a two-electron interchain transfer $t_{\perp}$,
we have the effective Hamiltonian for the charge sector of the
two chain system as
\be
H = -\sum_{j,a} [2t\tau_{j,a}^{z}
                + (V-J/4-2P)~\tau_{j,a}^{x}\tau_{j+1,a}^{x}]
-\sum_{j,a,b\neq a} [U_{\perp}\tau_{j,a}^{z}\tau_{j,b}^{z}
+ ~t_{\perp}~ (\tau_{j,a}^{x}\tau_{j,b}^{x}
+ \tau_{j,a}^{y}\tau_{j,b}^{y})]~,
\label{twochain}
\ee
where $a,b=1,2$ is the chain index. Denote the in-chain pseudospin
coupling as $J = (V-J/4-2P)$ and the inter-chain pseudospin coupling
as $J_{\perp} = U_{\perp}$.
Here, we study the strong coupling version of this problem in two limits (see below).
The weak-coupling problem is studied elsewhere~\cite{next}.

For the case of $|J_{\perp}| >> |J|,~ t_{\perp}$~, the 2 chain system can be better
thought of as strongly-coupled rungs which are weakly coupled to their
neighboring rungs. Thus, we treat $J$ as a 
weak perturbation on the zeroth-order system of rungs defined by the large
coupling $J_{\perp}$, giving $H_{eff} = H_{0} + H_{1}$ where
\beq
H_{0} &=& -h\sum_{j,a} \tau_{j,a}^{z}
+ J_{\perp}\sum_{j,a,b\neq a}\tau_{j,a}^{z}\tau_{j,b}^{z}\nonumber\\*
H_{1} &=& - J\sum_{j,a} \tau_{j,a}^{x}\tau_{j+1,a}^{x}
- \frac{t_{\perp}}{2}\hspace*{-0.2cm}\sum_{j,a,b\neq a}\hspace*{-0.2cm}(\tau_{j,a}^{+}\tau_{j,b}^{-}
+ {\rm h.c})
\label{start}
\eeq
where the effective magnetic field is given by $h=2t>0$.
\par
{\it LEH for $J_{\perp}<0$~.}
For $J_{\perp} <0$ and
$h<< J_{\perp}$, we find that the triplet state
$|+\rangle = \frac{1}{\sqrt{2}}(|\uparrow\downarrow\rangle +
|\downarrow\uparrow\rangle)$ and the singlet state $|-\rangle = 
\frac{1}{\sqrt{2}}(|\uparrow\downarrow\rangle - 
|\downarrow\uparrow\rangle)$ are degenerate on any rung and 
are separated from 
all other states by a large gap of order $\tJp$. Thus, these two states 
define the subspace which will determine the low-energy physics
of the system.
Identifying a pseudospin-1/2 operator $\xi_{j}$ with the low-energy
subspace on each rung, we treat the Hamiltonian $H_{1}$ as a
perturbation (to second order in $\tJ/\tJp$) to obtain the LEH as
\beq
H &=& \sum_{j}[-\frac{J^{2}}{2J_{\perp}}
(\frac{J_{\perp}^{2} - 2h^{2}}{J_{\perp}^{2} - 4h^{2}})
\xi_{j}^{z}\xi_{j+1}^{z} - \frac{t_{\perp}}{2}\sum_{j}\xi_{j}^{z}\nonumber\\*
&&\hspace*{-1.5cm} - \frac{J^{2}}{2J_{\perp}}(\frac{h^{2}}{J_{\perp}^{2}-4h^{2}})
(\xi_{j}^{+}\xi_{j+1}^{+} + {\rm h.c})
- \frac{J^{2}}{8J_{\perp}}(\frac{J_{\perp}^{2} - 2h^{2}}{J_{\perp}^{2} - 4h^{2}})]
\label{end}
\eeq
We find that $t_{\perp}$  acts as the strength of a Zeeman-splitting like term
in the LEH. Bosonising this, we obtain a sine-Gordon Hamiltonian with a 
cosine potential in the dual ($\theta$) field and a magnetic-field term  
\beq 
H=\frac{v}{2}[(\partial_{x}\phi)^{2} + (\partial_{x}\theta)^{2}] 
-\frac{m}{2\pi\alpha}\cos\beta_{1}\theta 
- \frac{\beta_{1} t_{\perp}}{2}\partial_{x}\phi
\eeq
We note that bosonisation of the general XYZ 
Hamiltonian results in the appearance of an additional $4k_{F}$ Umklapp term
~\cite{giamarchi}, $\cos\beta_{2}\phi$, which is irrelevant for a finite 
$t_{\perp}$ and is hence ignored in what follows. When $t_{\perp}$ is 
below a certain critical value, incommensurate Wigner charge order 
(ordering of the $\xi^{z}$) occurs~\cite{giamschulz}. Above this 
critical value, a spin-flop transition orders the system in the 
$x$ direction (i.e., ordering 
of the $\xi^{x}$) via a Kosterlitz-Thouless transition.
For $\beta_{1}^{2}<8\pi$, the cosine in the dual field is a relevant 
perturbation and orders the dual field. 
The magnetic-field term $\propto t_{\perp}$ leads  
to a ground state with charges which are coherently delocalised on the 
diagonals of each pair of nearest-neighbor rungs; this is an orbital 
antiferromagnet-type ordering with circulating currents in plaquettes
~\cite{tsvelik,giamarchi}.
\par
{\it LEH for $J_{\perp} > 0$~.}
For $J_{\perp}>0$,
and $h>0$, the triplet state
$|+\rangle = |\uparrow\uparrow\rangle$ is the low energy state
on any rung.  For $h=0$, we find that the triplet states
$|+\rangle$ (defined above) and
$|-\rangle = |\downarrow\downarrow\rangle$ are degenerate. Thus,
we can again identify these two states as the subspace which
determines the low-energy physics of the system.
For $h<<J$, we again identify a
pseudospin-1/2 operator $\xi_{j}$ with the low-energy
subspace on each rung, and treat the Hamiltonian $H_{1}$ as a
perturbation (to second order in $J/J_{\perp}$) to obtain the LEH as
\beq
H = -\frac{J^{2}}{4J_{\perp}}\sum_{j}\xi_{j}^{x}\xi_{j+1}^{x}- 2t\sum_{j}\xi_{j}^{z}~.
\eeq
This is just the 1D TFIM (with ferromagnetic Ising coupling).
In the ordered phase, the ground state has in-chain Wigner CO and dimers on 
every alternate rung.  The disordered phase is a gapped, short-ranged charge-dimer liquid.  At $t=J^{2}/4J_{\perp}$, the quantum critical point describes a gapless charge-dimer liquid with $\omega/T,vk/T$ QC scaling, exactly as was
described before.  Transposing the results obtained before, we conclude that the dc resistivity, optical conductivity, electronic Raman and dielectric responses
will be exactly described by the same scaling functions (eqs.(6)-(9)) with the 
gap, $\Delta_{\tau}$,  now being the CO gap of the ladder problem 
($H$ in eq.(15)).
Very interestingly, exactly such behavior is observed in undoped ladder system
$Sr_{14}Cu_{24}O_{41}$~\cite{blum} and attributed to a longitudinal, collective charge fluctuation mode, exactly as described here.
\par
{\it LEH for hole-doped ladder.} Upon doping the ladder with holes,
while a single hole experiences a linear confining potential in the
Wigner (Ising-like) or Peierls (dimerized) CO background, a pair of
holes on the same rung is free to propagate. One can then describe the
hole-pair as a hard-core boson, representing its creation and annihilation
operators using the spin-1/2 operators $\sigma^{\pm}$; the local charge
density is then described by $\sigma^{z}$. Following~\cite{tsvelik}, we
find the
LEH describing the dynamics of such hole-pairs to be the XXZ model
in an external magnetic field
\beq
H=\sum_{j}[-\frac{t_{\mathrm{h}}}{2}(\sigma_{j}^{+}\sigma_{j+1}^{-} + {\rm h.c})
- u_{\mathrm{h}}\sigma_{j}^{z}\sigma_{j+1}^{z} - \mu\sigma_{j}^{z}]
\eeq
where $t_{\mathrm{h}} \sim \tJ^{2}/\tJp$ is the pair-hopping matrix element,
$u_{\mathrm{h}}$ is the Coulomb interaction between pairs on nearest-neighbour
rungs and $\mu$ is the chemical potential of the holes. The phase
diagram of this model is known~\cite{tsvelik}: for $\mu=0$ and
$u_{\mathrm{h}}>t_{\mathrm{h}}$, the ground state is an insulating CDW of hole pairs.
Beyond a critical $\mu_{\mathrm{c}}=f(u_{\mathrm{h}},t_{\mathrm{h}})$, the system has a ground
state described by Bose condensation of hole pairs. In fact, from
the bosonisation analysis of the equivalent $S=1/2$ XXZ model in an
external Zeeman field, we know that
$<\sigma_{i}^{z}\sigma_{i+r}^{z}> \simeq r^{-1/\alpha}$ and
$<\sigma_{i}^{+}\sigma_{i+r}^{-}> \simeq r^{-\alpha}$ where
$\alpha=1/2-\pi^{-1}\sin^{-1}(2u_{\mathrm{h}}/t_{\mathrm{h}})$. Clearly, for $\alpha<1$,
the ground state has dominant superconducting correlations. This is
true for {\it both} the cases described above: in the first case, we
have a Bose condensate of intrachain pairs of holes, while in the
second hole pairs on individual rungs Bose condense, describing two
possible superconducting types in the ladder system. This finding
matches our conclusions obtained from a weak coupling analysis
~\cite{next}, and
thus constitutes a generic feature of undoped/doped strongly correlated
ladder systems.

To conclude, we have explored the strong-coupling limit of 
strongly correlated single chain and two-leg ladder models using 
a variety of methods.  Our results strongly suggest that prototype 
examples like organics (TMTSF) and $Sr_{14}Cu_{24}O_{41}$ lie in 
close proximity to an underlying QCP associated with charge order, 
and constitutes an advance in our understanding of their physical 
responses in a new theoretical framework.

\acknowledgments
We thank E. M\"{u}ller-Hartmann and G. I. Japaridze for several fruitful 
discussions.


\begin{thebibliography}{0}


\bibitem{org}
  \Name{McCarron III E. M. {\it et al.}}
  \REVIEW{Mater. Res. Bull.}{23}{1988}{1355};
  \Name{Uehara M. {\it et al.}}
  \REVIEW{J. Phys. Soc. Jpn.}{65}{1996}{2764};
  \Name{Jerome D.}
  \Book{Organic Superconductors: From $(TMTSF)_2PF_6$ to Fullerenes}
  \Publ{Marcel Dekker, New York}
  \Year{1994}
  \Page{405}.

\bibitem{kats}
  \Name{Zhuravlev A. K. and Katsnelson M. I.}
  \REVIEW{Phys. Rev. B}{64}{2001}{033102}.

\bibitem{tsvelik}
  \Name{Gogolin A. O., Nersesyan A. A. and Tsvelik A. M.}
  \Book{Bosonization and Strongly Correlated Systems}
  \Publ{Cambridge University Press, Cambridge}
  \Year{1998} and references therein. 

\bibitem{ogata} 
  \Name{Ogata M. and Shiba H.}
  \REVIEW{Phys. Rev. B}{41}{1990}{2326}.

\bibitem{haldane}
  \Name{Haldane F. D. M.}
  \REVIEW{Phys. Rev. B}{25}{1982}{4925}.

\bibitem{sachdev}
  \Name{Sachdev S.}
  \Book{Quantum Phase Transitions}
  \Publ{Cambridge University Press, Cambridge}
  \Year{1999} and references therein.

\bibitem{bra}
  \Name{Brazovskii S.}
  \REVIEW{cond-mat/0401309 preprint}{}{2004}{} and references therein;
  \Name{Staresinic D. {\it et al.}}
  \REVIEW{cond-mat/0509146 preprint}{}{2005}{}.

\bibitem{mon}
  \Name{Brazovskii S., Monceau P. and Nad F.}
  \REVIEW{Synthetic Materials}{137}{2003}{1331}. 

\bibitem{bishop}
  \Name{Fehske H. {\it et al.}}
  \REVIEW{Physica B}{359-361}{2005}{699}.

\bibitem{next}
  \Name{Lal S. and Laad M. S.}
  \REVIEW{to be submitted}{}{2005}{}.

\bibitem{giamarchi}
  \Name{Giamarchi T.}
  \Book{Quantum Physics in One Dimension}
  \Publ{Oxford University Press, Oxford}
  \Year{2004} and references therein.

\bibitem{giamschulz}
  \Name{Giamarchi T. and Schulz H. J.}
  \REVIEW{J. de Physique (Paris)}{49}{1988}{819}.

\bibitem{blum}
  \Name{Blumberg G. {\it et al.}} 
  \REVIEW{Science}{297}{2002}{584}.



\end{thebibliography}
\end{document}